\documentclass[12pt]{iopart}
\usepackage{dsfont}    % for `\realline' and `\complexnumb'
\usepackage{graphicx}  % for plots, `\rotatebox' etc.
\usepackage[T1]{url}   % urls (in bibliography) with T1-Encoding

\providecommand{\realline}{\mathds{R}}
\providecommand{\complexnumb}{\mathds{C}}
\providecommand{\sunmass}{\mathrm{M}_{\scriptscriptstyle\odot}}
\providecommand{\prob}{\mathrm{P}}
\providecommand{\imag}{\mathrm{i}}
\providecommand{\differential}{\mathrm{d}}
\providecommand{\expectation}{\mathrm{E}}
\providecommand{\newton}{\mathrm{G}}
\providecommand{\lightspeed}{\mathrm{c}}

\begin{document}
\urlstyle{sf}
 \title[Bayesian inference on compact binary inspiral signals...]
       {Bayesian inference on compact binary inspiral 
        gravitational radiation signals in interferometric data}
 \author{Christian R\"{o}ver$^1$,
         Renate Meyer$^1$ and 
         Nelson Christensen$^2$}
\address{$^1$ Department of Statistics, The University of Auckland,
         Auckland, New Zealand}
\address{$^2$ Department of Physics \& Astronomy, Carleton College,
         Northfield, MN, USA}
\eads{\mailto{christian@stat.auckland.ac.nz}, 
      \mailto{meyer@stat.auckland.ac.nz},
      \mailto{nchriste@carleton.edu}}

\begin{abstract}
Presented is a description of a Bayesian analysis framework 
for use with interferometric gravitational radiation data 
in searches for binary neutron star inspiral signals. 
Five parameters are investigated, and the information extracted from the data 
is illustrated and quantified. 
The posterior integration is carried out using Markov chain 
Monte Carlo (MCMC) methods. 
Implementation details include the use of importance resampling 
for improved convergence and informative priors reflecting
the conditions expected for realistic measurements. 
An example is presented from an application using realistic, 
albeit fictitious, data.
We expect that these parameter estimation techniques will prove useful
at the end of a binary inspiral detection pipeline for interferometric
detectors like LIGO or Virgo.
\end{abstract}

%  04.80.Nn :  `Gravitational wave detectors and experiments'
%  02.70.Uu :  `Applications of Monte Carlo methods'

\submitto{Classical and Quantum Gravity}

\section{Introduction}
Great effort is currently being expended in the search for gravitational
radiation. Detectors around the world 
\cite{AbbottEtAl2004a,AcerneseEtAl2005,AndoEtAl2005,Hewitson2005}
are reaching sensitivities where an
event may be seen in the near future. The field of research has grown immensely
since the first prediction of gravitational radiation \cite{Einstein1916}, or even
from the time of the observations that confirmed its existence 
\cite{TaylorWeisberg1989}. 
The inspiral of compact binary objects may provide the cleanest
system for comparing observations with general relativistic predictions 
\cite{Thorne1987}.
Observation of inspiral events could provide important information on the
structure of neutron stars 
\cite{CutlerEtAl1993,Hughes2002}. 
Even cosmological information can be
extracted from the observation of inspiral events 
\cite{Schutz1986,Markovic1993,CutlerFlanagan1994,Finn1996}.
The characteristics of radiation in the post-Newtonian regime will provide insight
into highly non-linear general relativistic effects, such as the observation of
the formation of a Kerr black hole as the binary system decays 
\cite{CutlerFlanagan1994,FlanaganHughes1998a,FlanaganHughes1998b}.
LIGO has recently conducted searches for signals from the inspiral of binary
neutron star systems in science run data 
\cite{AbbottEtAl2004b, AbbottEtAl2005}.
\\ \indent
Detection pipelines for binary inspiral signals 
\cite{AbbottEtAl2004b,MarionEtAl2003,AmicoEtAl2003}
in interferometer
data will likely provide the first indication of detection. 
These pipelines provide parameter estimation 
through various estimation techniques.
Other techniques, like those that we will describe in this paper, 
provide a means to examine the data and produce not just point estimates 
of parameters, 
%but their complete distributions including summary statistics. 
but provide more detailed information 
e.g. on accuracy and correlations of parameters given the data at hand.
The unique analysis methods that we bring to the examination of
LIGO data are computationally intensive MCMC algorithms. Our method of
parameter estimation is an exercise in Bayesian inference. These Bayesian
statistical techniques offer great promise in problems where the number of
parameters is large. Since its initial application in digital signal analysis
\cite{GemanGeman1984}
MCMC methods have revolutionised many areas of applied statistics
\cite{Loredo1992}. 
A distinct advantage of the MCMC approach is that computational time does not
grow exponentially with parameter number, as it does for other methods 
\cite{MCMCinPractice}.
After their introduction to the cosmological parameter estimation problem
\cite{ChristensenMeyer2000,ChristensenMeyerKnoxLuey2001}, 
MCMC methods were quickly adopted and widely used with CMB data
\cite{KnoxChristensenSkordis2001,RubinoEtAl2003,DunkleyEtAl2005}, 
including WMAP 
\cite{SpergelEtAl2003,VerdeEtAl2003}.
We anticipate a similar scenario with gravitational radiation; after initial
detection attention will be focused on parameter estimation and associated
astrophysical information extracted from the data. The techniques that we
describe in this paper are intended to be applied to interferometer data after
an initial event has been registered by an inspiral detection pipeline 
\cite{AbbottEtAl2004b,MarionEtAl2003,AmicoEtAl2003};
our MCMC method would allow the extraction of information about the physical
parameters.
The technique we present here provides parameter estimates from the data output
of a single interferometer. An MCMC technique using the coherent addition 
of signals from multiple detectors will be presented in a future publication.
\\ \indent
In the present paper we describe the MCMC routine we have developed to examine
interferometer data for binary neutron star inspiral signals. 
We have implemented advanced
MCMC methods in order to effectively and efficiently search a five-parameter
space, 
%find the signal, and produce summary statistics. 
find the signal's parameter values, and quantify and characterise them in detail.
There have been significant advances in the execution of the binary neutron star 
inspiral MCMC routine since the publication of our previous method 
\cite{ChristensenMeyerLibson2004};
presented here is a description of our routine, implemented in~C,
that employs advanced MCMC techniques in order 
to increase efficiency and performance.
\\ \indent
The outline of the paper is as follows. Section~\ref{sec:AnalysisStrategy} 
describes our analysis
strategy and contains a description of Bayesian parameter estimation methods,
and the techniques we employ in the binary neutron star inspiral MCMC application. 
In Section~\ref{sec:InferenceExample}
we present an example application of our method using fictitious data; 
a binary neutron star
inspiral event is embedded within noise (that matches the LIGO design
sensitivity). We conclude the paper with a summary of our results, and a
description of our intentions to expand this work to more complex situations. 

\section{Analysis strategy} \label{sec:AnalysisStrategy}
\subsection{Bayesian modeling and MCMC posterior simulation}
%We begin with 
Presented here is
a short introduction to Bayesian inference, 
which is quite a different way to approach problems like parameter 
estimation compared to more commonly used ``frequentist'' methods,
%like e.g.\ 
such as
maximum likelihood estimation.
The model setup in\-cludes the specification of a \emph{prior distribution} 
for all involved parameters as well as the \emph{sampling distribution} 
for observed quantities. 
Inference is done via the derived \emph{pos\-te\-ri\-or distribution} given the 
observed data; 
this allows one to make probabilistic statements
about parameters of interest by treating them as random variables. 
Due to its setup the concept is sometimes dubbed 
\emph{full probability modeling}. 
While the theory behind Bayesian modeling is straightforward, 
its practical application for inference often requires 
the use of numerical methods.
\\ \indent
The \emph{sampling distribution} is specified in terms of the corresponding
probability density function (PDF) $p(y|\theta)$, defining how the observations~$y$ 
come about given a (fixed) parameter setting~$\theta$.
Viewed as a function of~$\theta$, it also defines the \emph{likelihood function}.
%The use of the likelihood function,which may be thought of it as `linking 
%observations to parameters', is also common in frequentist analyses.
Many frequentist techniques, such as maximum likelihood estimation, 
are also based on the likelihood function, which may be thought of it as `linking 
observations to parameters'.
\\ \indent
In the frequentist approach $\theta$~is viewed as fixed but unknown. 
%With a Bayesian approach one treats~$\theta$ as
In the Bayesian paradigm, on the other hand, $\theta$ is treated as
a random variable with a probability
distribution that reflects the re\-searcher's uncertainty about the parameters.
In addition to the likelihood a prior PDF~$p(\theta)$ is used, 
reflecting the pre-experimental knowledge and uncertainty about the parameters. 
The availability of both prior PDF~$p(\theta)$ and likelihood $p(y|\theta)$ 
then allows to derive the \emph{posterior PDF} of~$\theta$ 
conditional on the observed data~$y$
\begin{equation} \label{eqn:BayesTheorem}
  p(\theta|y) \;=\; \frac{p(\theta)\, p(y|\theta)}{p(y)} \;\propto\; p(\theta)\, p(y|\theta)
\end{equation}
by applying 
%(a generalisation of) 
Bayes' theorem
%In general, Bayes' theorem deals with conditional probabilities, here it provides
%the probability density of~$\theta$ \emph{conditional on the observed data $y$}
\cite{BDA}.
Note that the denominator
$p(y)=\int p(y,\theta) \differential\theta = \int p(y|\theta)p(\theta)\differential\theta$ 
is a normalising constant that does not depend on~$\theta$, 
so the posterior PDF~$p(\theta|y)$ is essentially 
proportional to the product of prior PDF and likelihood function.
\\ \indent
The posterior PDF expresses the knowledge about the parameter(s) 
given model (prior PDF and likelihood function) \emph{and} data.
Inference then aims at deriving probabilistic statements about parameters 
(or related quantities) from the posterior; one might for example be interested in
posterior expectations, quantiles or marginal PDFs of quantities of interest, 
or probabilities of certain events. 
\\ \indent
Note that the impact of the prior specification is limited, since 
asymptotic theory shows that the importance of the prior disribution 
diminishes as the sample size increases, so the posterior is eventually dominated 
by the likelihood \cite{BDA,LeCam1953}.
When MCMC methods are used for posterior analysis as in this paper, 
the results (MCMC samples) yielded using a certain prior 
may easily be recycled to generate posterior samples assuming different priors
by using importance sampling 
\cite{BDA}.
% BDA, section 10.3, page 307 sqq
\\ \indent
\emph{Frequentist} and \emph{Bayesian} analyses are fundamentally different
in their approach to given problems; Finn \cite{Finn1997} puts the difference as
``guessing nature's state'' in case of frequentist analysis as opposed to
``learning from observation'' for Bayesian analysis.
%For more detailed discussions see e.g.~\cite{Loredo1992,BDA,Finn1997}.
For more detailed discussions see e.g.~\cite{Loredo1992,Finn1997,Jaynes1986}.
\\ \indent
For posterior computations we use a simulation-based approach as the posterior 
distribution is high-dimensional and the calculation of marginal summaries
like marginal posterior means would require high-dimensional integration. 
The simulation-based approach is to
generate a sample from the posterior 
and then approximate the desired integrals by sample averages. 
%For example, for some random variable~$X$ with probability density~$p_X(x)$ 
%its expectation $\expectation [X] = \int x\,p_X(x)\,\differential x$ can be 
%approximated through a sample average $\bar{x} = \frac{1}{N} \sum_{i=1}^N x_i$ 
%where $x_1,\ldots,x_N$ are random draws from $p_X(x)$.
For example, the expectation of a random variable~$X$ with PDF~$p_X(x)$, 
$\expectation [X] = \int x\,p_X(x)\,\differential x$ can be 
approximated through a sample average $\bar{x} = \frac{1}{N} \sum_{i=1}^N x_i$ 
where $x_1,\ldots,x_N$ are random draws from $p_X(x)$.
Analogously, marginal PDFs, quantiles etc.\ can be estimated from samples 
from the distribution of interest.
\\ \indent
Markov chain Monte Carlo (MCMC) methods in general may be used to generate 
random sequences of numbers that have a specified stationary distribution 
and in which each random step in the sequence only depends on the previous stage 
of the chain. 
%The Metropolis-algorithm in particular can be applied if the target distribution 
%is specified in terms of its density which only needs to be known
The Metropolis-algorithm in particular can be applied if the target PDF is known only
up to a normalising constant.
  Note that this is the case in equation~(\ref{eqn:BayesTheorem}), as the
  evaluation of the numerator~$p(y)$ would involve high-dimensional integration.
While a Metropolis sampler is guaranteed to function as the number of samples goes 
towards infinity, its proper behaviour and efficiency within finite time heavily 
depends on the sensible specifications of its \emph{starting point} and 
\emph{proposal distribution} 
\cite{BDA}.
% BDA, section 11.5, page 333

\subsection{Likelihood, signal templates \& priors}
\begin{figure}[ht]
  \begin{center}
    \includegraphics[width=12cm]{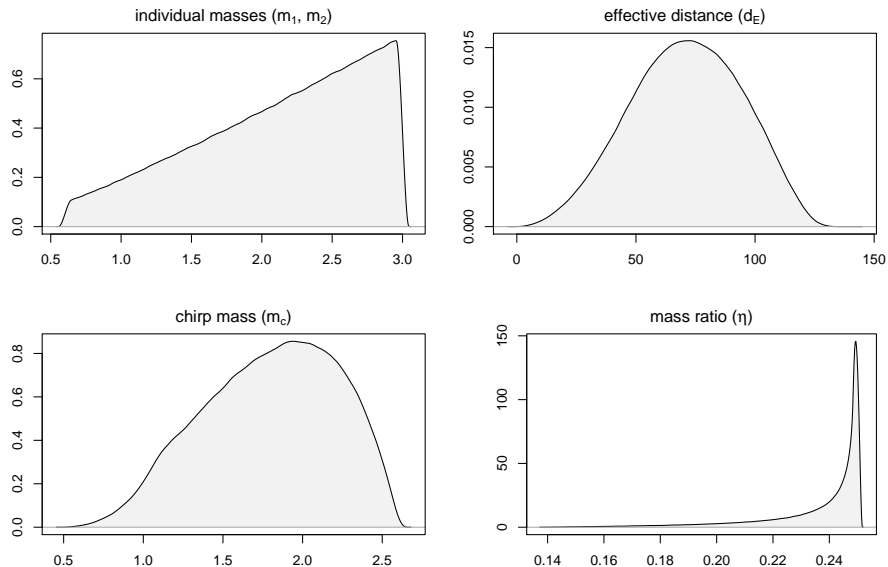}
    \caption{Marginal \emph{prior} PDFs for masses 
             and effective distance.
             The upper left plot shows the PDF of individual masses ($m_1$, $m_2$), 
             the lower plots show PDFs in the alternative 
             mass parametrisation of chirp mass~($m_c$) and mass ratio~($\eta$).}
    \label{gra:Priors}
  \end{center}
\end{figure}

The observed data are a (real-valued) time series of length~$N$ and 
sampling rate~$\frac{1}{\Delta_t}$ indicating the phase shift between the 
two interferometer arms over time at evenly spaced time points.
In order to analyse a data set, it is Fourier-transformed, from
\begin{equation}
  \{z(t)\in\realline:\; t=0,\Delta_t,2\Delta_t,\ldots,(N-1)\Delta_t\}
\end{equation}
into
\begin{equation} \textstyle
  \{\tilde{z}(f)\in\complexnumb:\; f=0,\Delta_f,2\Delta_f,\ldots,(\frac{N}{2}-1)\Delta_f\},
\end{equation}
where $\Delta_f = \frac{1}{N\Delta_t}$ and
\begin{equation}
  \tilde{z}(f) = \Delta_t \sum_{j=0}^{N-1} z(j\Delta_t) \exp(-2 \pi \imag j f).
\end{equation}
The likelihood function depends on the data and a \emph{signal template}, describing 
the theo\-reti\-cal\-ly derived detector response~$s_\vartheta$ for a given 
parameter set $\vartheta$. 
%\\ \indent
The likelihood of some parameter set~$\vartheta$ then is proportional to the sum 
of the squared and normalised differences between Fourier transforms of observed 
signal~($\tilde{z}$) and signal template~($\tilde{s}_\vartheta$) over the discrete 
set of Fourier frequencies  $\{(j\times\Delta_f):\;j_L\leq j\leq j_U\}$ 
\cite{FinnChernoff1993,ChristensenMeyer2001}:
\begin{equation}
  \label{eqn:chisquare}
  p(z|\vartheta) = K\times\exp\Biggl(-\frac{2}{N\Delta_t}\sum_{j=j_{L}}^{j_{U}}
  \frac{|\overbrace{\tilde{z}(j\times\Delta_f)}^{\mbox{data}}-
         \overbrace{\tilde{s}_\vartheta(j\times\Delta_f)}^{\mbox{template}}|^2}
       {\underbrace{S_n(j\times\Delta_f)}_{\mbox{noise PSD}}}
  \Biggr)
\end{equation}
where $j_L\times\Delta_f$ and $j_U\times\Delta_f$ are the lower and upper bounds 
of the examined frequency range, 
% $\Delta_f$ is the resolution of the (discrete) Fourier transformed data, 
%$|\cdot|$ denotes the absolute value of the (complex-valued) difference, 
$S_n(\cdot)$ is the (estimated) noise power spectral density (PSD), 
$N\Delta_t$ is the length of the analysed data segment, 
and $K$ is a normalising constant.
\\ \indent
The signal templates we use are 2.0-post-Newtonian stationary phase 
approximations of the Fourier-transformed inspiral waveform \cite{TanakaTagoshi2000}.
The model has five parameters:
the individual masses of the two involved companions $m_1$ and $m_2$,
the coalescence time~$t_c$,
the coalescence phase~$\phi_0$ and
the effective distance~$d_E$.
The effective distance is not the actual distance to the source, but also reflects
the effect on the amplitude of the gravity wave from other parameters 
and is in general greater than the actual distance.
The other parameters affecting the gravity wave amplitude are the binary
inspiral system's sky position and orbital plane inclination angle, the
interferometer orientation, and the polarization of the wave.  We also ignore
any effects due to the spins of the compact objects.
The template is then defined in terms of 
\emph{total mass} $m_t=m_1+m_2$ and \emph{mass ratio} $\eta = \frac{m_1m_2}{m_t^2}$ as:
\begin{equation}\label{eqn:cosinechirp}
  \tilde{s}_\vartheta(f) = 
  \frac{\sqrt{\eta}\;m_t^{\frac{5}{6}}}{d_E} \;
  \frac{\sqrt{5}\;\newton^{\frac{5}{6}}}{2\sqrt{6}\;\pi^{\frac{2}{3}}\;\lightspeed^{\frac{3}{2}}} \;
  f^{-\frac{7}{6}} \; 
  \exp\Bigl(-\imag\bigl(\underbrace{\psi(f) + \phi_0 + 2\pi f t_c}_{\mbox{phase evolution}}\bigr)\Bigr)
\end{equation}
where  
%$\newton=6.672\,59 \times 10^{-11} \frac{\mathrm{m}^3}{\mathrm{kg}\;\mathrm{s}^2}$ 
%is the \emph{Gravitational} or \emph{Newton's constant}, 
%$\lightspeed=299\,792\,458 \frac{\mathrm{m}}{\mathrm{s}}$ is the speed of light, 
%and
\numparts
\begin{equation}
  \label{eqn:psifun}
  \psi(f) := \sum_{i=1}^{4}a_i\zeta_i(f),
\end{equation}
\begin{eqnarray}
  a_1 &=& \frac{3}{128\eta} q^{-\frac{5}{3}},\\
  a_2 &=& \frac{1}{384\eta} \left(\frac{3715}{84}+55\eta\right)q^{-1},\\
  a_3 &=& - \frac{1}{128\eta} 48\pi q^{-\frac{2}{3}},\\
  a_4 &=& \frac{3}{128\eta} 
          \left(\frac{15\,293\,365}{508\,032}+\frac{27\,145}{504}\eta+\frac{3085}{72}\eta^2\right) 
          q^{-\frac{1}{3}},
\end{eqnarray}
\endnumparts
$\zeta_1(f)=f^{-\frac{5}{3}}$, 
$\zeta_2(f)=f^{-1}$,
$\zeta_3(f)=f^{-\frac{2}{3}}$,
$\zeta_4(f)=f^{-\frac{1}{3}}$,
and $q=\pi \newton m_t \lightspeed^{-3}$ 
\cite{ChristensenMeyer2001,TanakaTagoshi2000}.
\\ \indent
Priors are specified with respect to some preliminary considerations.
The co\-a\-les\-cence time~$t_c$ is assumed to be known in advance with a certain 
accuracy through preprocessing of the data
\cite{AbbottEtAl2004b,MarionEtAl2003,AmicoEtAl2003}; 
we set its prior to be uniform across $\pm 5 \mbox{ms}$ around the true value 
(which of course is known for our simulated data).
The prior for the coalescence phase~$\phi_0$ is uniform across its domain $[0,2\pi]$.
\\ \indent
The prior for masses $m_1$, $m_2$ and effective distance $d_E$ is set 
in order to reflect the distribution of parameters of a neutron star inspiral
\emph{given} that it has been detected in the first place.
Initially, the prior for two companions' individual masses ($m_1$ and $m_2$) 
is uniform between~0.6 and 3.0~$\sunmass$ 
%(solar masses: $\sunmass=1.988\,919\times10^{30}\;\mbox{kg}$), 
(solar masses: $\sunmass\approx 2 \times10^{30}\;\mbox{kg}$), 
which effectively covers the range of values expected for binary neutron star
systems.
The prior for $d_E$ is derived from the assumption that inspirals happen 
uniformly across space, so that $\prob(d_E\leq x) \propto x^3$. 
So far, this leads to an improper PDF (that has an infinite integral).
For actual data, in order to be investigated at all, an inspiral event needs to
emit a signal that 
is strong enough to produce a trigger in an inspiral search routine
\cite{AbbottEtAl2004b,MarionEtAl2003,AmicoEtAl2003},
and thus cannot originate from arbitrarily great distances.
We incorporated this constraint by downweighing the 
low-mass/great-distance
inspirals according to their lower \emph{detection probability}.
The probability of detection is assumed to depend simply on the signal's amplitude, 
which is affected by the inspiral's masses and distance and is proportional to:
\begin{eqnarray} \label{eqn:alpha}
  \alpha(m_1, m_2, d_E)
      & = &  {\frac{\scriptstyle 1}{\scriptstyle 2}}(\log(m_1)+\log(m_2)) - {\frac{\scriptstyle 1}{\scriptstyle 6}} \log(m_1+m_2) - \log(d_E)\\
      &  = & \log\Biggl(\frac{\sqrt{m_1 m_2}}{(m_1+m_2)^{\frac{1}{6}}\,d_E}\Biggr)
         \; = \; \log\Biggl(\frac{\sqrt{\eta}\;m_t^{\frac{5}{6}}}{d_E}\Biggr) \nonumber
\end{eqnarray}
(so $\alpha$ denotes the logarithmic amplitude; see also equation~(\ref{eqn:cosinechirp})).
We could have set a threshold amplitude below which neutron star inspirals would be assumed 
to be undetectable, but favoured a `smoother' transition that does not explicitly 
assign zero probability to parts of the parameter space.
We model the dependence between a given (logarithmic) amplitude~$x$ 
and detection probability using a (sigmoidal) logistic function of the form:
\begin{equation} \label{eqn:SigmoidFunction}
  d_{a,b}(x) = \frac{1}{1+\exp(\frac{x-a}{b})}
\end{equation}
where $a$ and $b$ are set so that $d_{a,b}(x_U)=1-p$ and $d_{a,b}(x_L)=p$ 
for some $0<p<0.5$ (e.g.:~$p:=0.1$) and upper and lower reference points 
$x_U$ and $x_L$. 
So $x_U$ denotes the amplitude at which the detection probability reaches~$(1-p)$, 
and $x_L$ is the amplitude where the probability falls below~$p$. 
In order to fit~$d$ through these points, its parameters are set to:
\begin{equation}
  \label{eqn:SigmoidParameters}
  a:= \frac{x_L+x_U}{2} \qquad \mbox{and} 
  \qquad 
  b:= \frac{x_U-x_L}{2\,\log(\frac{p}{1-p})}.
\end{equation}
So, given the above prior \emph{occurrence} and \emph{detection} probabilities, 
the resulting (proper) joint prior PDF for individual masses~$m_1$ and~$m_2$ 
and distance~$d_E$ is:
\begin{equation}
  p(m_1, m_2, d_E) \propto \mathrm{I}_{[0.6,3.0]}(m_1) \times \mathrm{I}_{[0.6,3.0]}(m_2) 
                               \times d_E^2\times d_{a,b}(\alpha(m_1, m_2, d_E))
\end{equation}
where 
$\mathrm{I}_{\mathcal{A}}(x):=\left\{\begin{array}{ll} 1 & \mbox{if } x\in\mathcal{A} \\ 0&\mbox{otherwise}\end{array}\right.$ 
is the indicator function for any set~$\mathcal{A}$.
For the examples given in this paper we have chosen values for the sigmoidal
logistic function so that a \mbox{2-2$\sunmass$} inspiral would be detectable 
out to 90 and 95~Mpc with probabilities of 90\% and 10\% respectively. 
These values effectively produce a prior PDF for the effective 
distance that smoothly covers the values of interest, 
and then falls to zero as the distance gets too large. 
\\ \indent
Fig.~\ref{gra:Priors} shows estimates
of the marginal prior PDFs of masses and effective distance, 
in terms of individual masses ($m_1$, $m_2$) 
as well as in the alternative parametrisation 
%of chirp mass~($m_c$) and mass ratio~($\eta$).
of \emph{chirp mass}~($m_c = \frac{(m_1 m_2)^{0.6}}{(m_1+m_2)^{0.2}}$) 
and \emph{mass ratio}~($\eta = \frac{m_1 m_2}{(m_1+m_2)^2}$).
Greater individual masses are more likely to be observed 
since low-mass inspirals need to be close to be detectable, 
while high-mass inspirals may as well originate from greater distances: 
a feature that is also known as the \emph{Malmquist effect} \cite{Sandage2001}.
Note that the marginal prior PDF for $d_E$ extends well beyond 95~Mpc, 
because the thresholds $x_L$ and $x_U$ were specified for 
\mbox{2-2$\sunmass$} binary neutron star inspirals;
inspirals involving greater masses still generate signals of observable 
amplitudes when happening further out.

\subsection{Technical details}
The MCMC sampler is implemented in~C\@. 
Data is imported from the \emph{Frame File format} using the \emph{Frame Library} 
\cite{FrameLibrary}. 
Before transformation, the data are low-pass-filtered and downsampled 
(from $16\,384$ to $4096$~Hz)
\cite{IEEE-DSP-8.2}; 
required Filter coefficients are determined by the `Parks-McClellan'-algorithm 
(see e.g.~\cite{DSP}), again using existing software \cite{ParksMcClellan}. 
The Discrete Fourier Transform of the data (for both likelihood computation 
and spectral density estimation) is carried out using the free 
\emph{FFTW} library \cite{FFTW}.
The noise spectrum is estimated from a data section that is 
disjoint from the actually investigated data \cite{Welch1967}. 
\mbox{(Pseudo-)} random numbers are generated using the \emph{Randlib} 
library \cite{randlib}.
\\ \indent
The frequency range over which the likelihood is computed 
(see equation~(\ref{eqn:chisquare})) was set to \mbox{40--1800~Hz}. 
In the original mass parametrisation of individual masses~$m_1$ and~$m_2$ 
the two masses showed high posterior correlation, 
making sampling from the posterior difficult. 
In order to improve sampling we expressed the likelihood in terms of 
chirp mass~($m_c$) and mass ratio~($\eta$) instead.
%where
%$m_c = \frac{(m_1 m_2)^{0.6}}{(m_1+m_2)^{0.2}}$ and
%$\eta = \frac{m_1 m_2}{(m_1+m_2)^2}$. 
In addition we reparameterised the effective distance from~$d_E$ to $\log(d_E)$,
thus implicitly yielding an unbounded parameter space and
proposal step widths that are proportional to the %actual 
distance~($d_E$) itself.
\\ \indent
Approximate posterior samples as starting values for the MCMC chains 
are generated using \emph{importance resampling},
by first generating a large sample from a distribution covering the whole prior,
and then drawing the actual sample out of these 
with correspondingly assigned weights depending on the posterior PDF
\cite{BDA,SmithGelfand1992}. % BDA, section 10.5, page 312 sqq.
The Metropolis sampler's proposal distribution was chosen to be 
a Multivariate \mbox{Student-t} distribution.
The t-distribution has a similar shape to the Normal dis\-tri\-bu\-tion,
but posesses an additional degrees-of-freedom parameter~$\nu$,
and for $\nu\rightarrow\infty$ approaches a Normal distribution.
It has `heavier tails' than a Normal distribution, which means that extreme 
values are more likely under a t-distribution than they were under a Normal, 
making it a more robust choice as a proposal distribution 
\cite{BDA}. % BDA, appendix A.2, page 481
The degrees of freedom were set to $\nu:=3$, the lowest possible integer 
value for which the distribution's variance is finite.
Starting off from an initial setting, the covariance parameter of the proposal 
distribution is recursively adapted to the sample covariance of generated samples 
during an initial burn-in phase \cite{Welford1962,Chan1983}.
The scale of the proposal covariance is set to $\frac{1}{10}$~of the sample 
covariance estimate, yielding a reasonable acceptance rate for the sampler.
\\ \indent
Convergence of the Markov chains is monitored by using several chains 
that are run simultaneously from different starting points,
so one may e.g.\ verify whether these eventually end up in the same mode
\cite{GelmanRubin1992}.
In order to minimise the effect of correlated draws from the MCMC output,
only every 50th MCMC draw is written to a file which is then imported 
into~R, a statistical software, for eventual analysis
\cite{R-Manual}.
The C~code generates some 2500~MCMC samples per minute 
on a 3.2~GHz Pentium~4 desktop PC.
%  2530.1 / min. 
% \\ \indent
The univariate marginal PDFs that are shown in this paper are 
\emph{kernel density estimates} 
(for more details about these see e.g.\ \cite{Scott}). 
% Epanechnikov kernel, Silverman bandwidth.
%The bivariate densities (Fig.~\ref{gra:BivariateDensities}) are 2-dimensional histograms, 
%the greyscale plots showing relative densities normalised to the mode;
%the actual density values are not relevant here.

\section{Example application with simulated data} 
\label{sec:InferenceExample}
\begin{figure}[ht]
  \begin{center}
    \includegraphics[width=15cm]{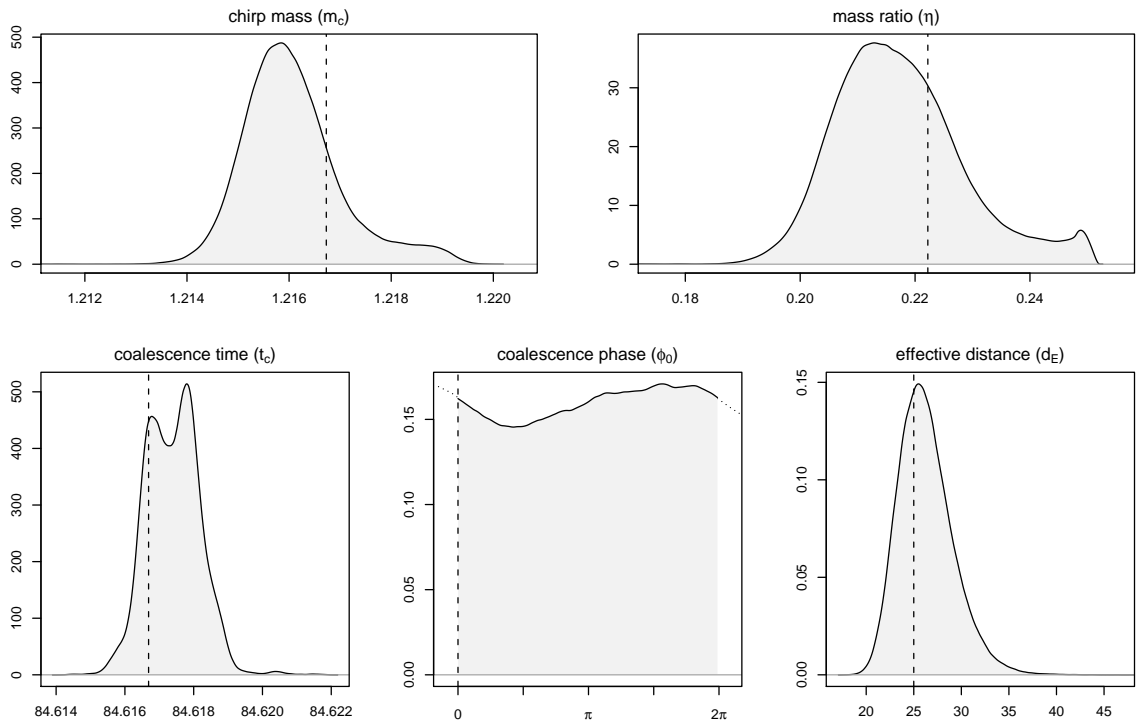}
    \caption{Marginal posterior PDFs of the five parameters.
             Dashed lines indicate true parameters.}
    \label{gra:MarginalDensities}
  \end{center}
\end{figure}

\begin{figure}[ht]
  \begin{center}
    \includegraphics[width=15cm]{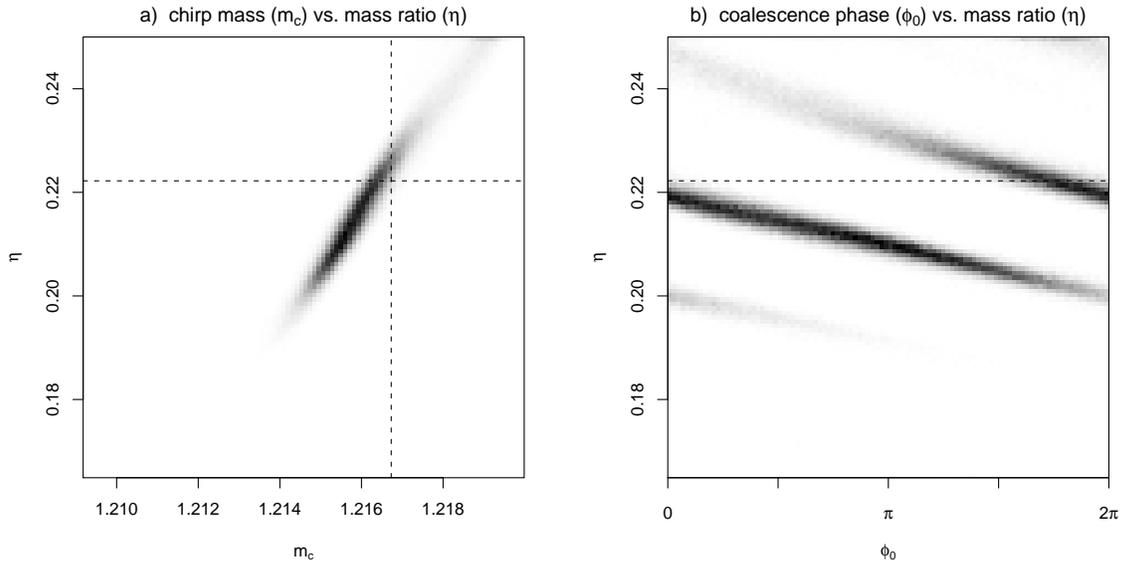}
    \caption{Bivariate marginal posterior PDFs for two pairs of parameters.
             Dashed lines indicate the true values,
             the true coalescence phase is~$\phi_0 = 0$ 
             (Histograms, the greyscale plots show relative PDFs
              normalised to the mode).}
    \label{gra:BivariateDensities}
  \end{center}
\end{figure}

\begin{figure}[ht]
  \begin{center}
    \includegraphics[width=12cm]{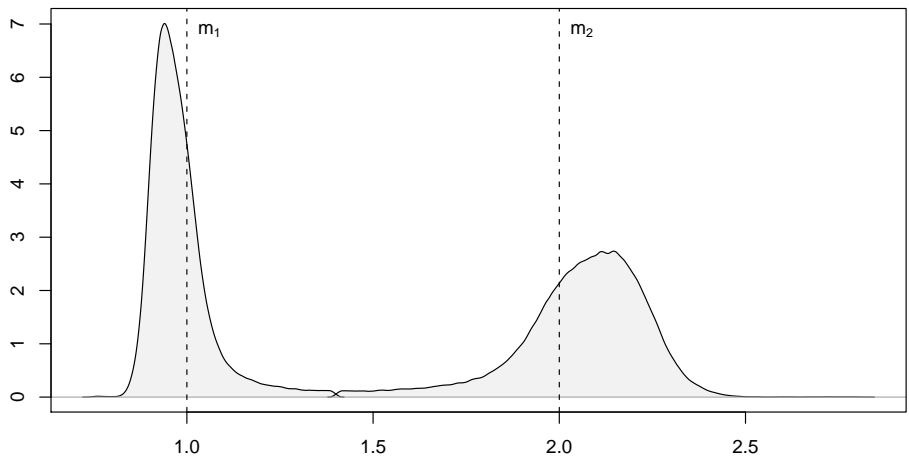}
    \caption{Combined plot of marginal posterior PDFs of the two companions' 
             in\-di\-vid\-u\-al masses (assuming $m_1 \leq m_2$). 
             Dashed lines indicate true parameters.}
    \label{gra:MassDensities}
  \end{center}
\end{figure}

\begin{figure}[ht]
  \begin{center}
    \includegraphics[width=12cm]{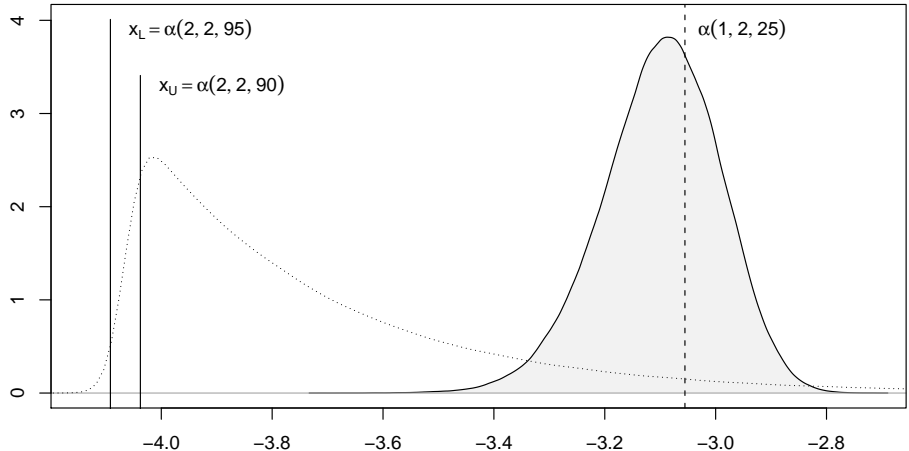}
    \caption{Posterior PDF of the signal's logarithmic am\-pli\-tude 
             $\alpha(m_1, m_2, d_E)$; the dashed line indicates the true value.
             The prior PDF (dotted line) and $x_U$ and 
             $x_L$ are shown as well.}
    \label{gra:AlphaDensity}
  \end{center}
\end{figure}

We illustrate the results of a run of the MCMC sampler on simulated data 
for which the true parameter values are known. 
 The signal analysed had an effective distance of 25~Mpc, and was
 embedded in Gaussian and stationary noise that had its noise power
 spectral density match that of LIGO's target sensitivity \cite{Sigg2004}.
 The embedded signal had a signal to noise ratio of~10.
Six parallel chains were run; the starting points of the chains were generated 
by importance resampling of $100\,000$ draws, a number that proved to yield 
enough eventual draws that were sufficiently close to the main posterior mode 
to ensure reliable and fast convergence of the Metropolis algorithm.
The first $30\,000$ iterations of each chain were considered the burn-in-phase, 
during which the iterations $15\,000$--$30\,000$ were used to tune the proposal 
covariance.
The code was then run for 2~million iterations in total, which after thinning 
out of the samples 
%(every 50th was kept) 
and discarding the burn-in
yielded a sample of size~\mbox{$236\,400$} from the posterior.
The `multivariate potential scale reduction factor' $\widehat{R}^p$ 
was close to~1 ($\widehat{R}^p = 1.0034$), 
% $mpsrf
% [1] 1.003391
indicating convergence of all chains \cite{BrooksGelman1998}.
\\ \indent
Figures~\ref{gra:MarginalDensities}--\ref{gra:AlphaDensity} 
illustrate estimates of \emph{marginal} posterior PDFs, 
that is, univariate PDFs with respect to one specific parameter, 
with all other parameters integrated out.
Firstly, Fig.~\ref{gra:MarginalDensities} shows posterior PDFs for 
the five individual parameters. 
Most of them exhibit a mode near the true parameter value (indicated by 
dashed lines). 
One can see that the relative precision of parameter estimation 
varies significantly between different parameters. 
For example, the posterior of the chirp mass covers a range of about 
0.006~$\sunmass$, while the prior range initially was some 2.1~$\sunmass$ 
(cp.\ Fig.~\ref{gra:Priors}).
The coalescence phase's posterior, on the other hand, 
still covers the complete prior domain.
\\ \indent
Fig.~\ref{gra:BivariateDensities} allows for some insight into
joint PDFs of some of the parameters.
The joint PDF of chirp mass and mass ratio (\ref{gra:BivariateDensities}a) 
shows a positive correlation between the two parameters.
Fig.~\ref{gra:BivariateDensities}b shows interaction 
between two parameters ($\phi_0$ and $\eta$), 
and in particular demonstrates that although the marginal PDF of~$\phi_0$ 
alone is almost uniform (see Fig.~\ref{gra:MarginalDensities}), 
this does not imply that its effect on the posterior was negligible. 
\\ \indent
The MCMC sampler internally works with chirp mass ($m_c$) and mass ratio ($\eta$) 
instead of individual masses ($m_1$, $m_2$).
A posterior sample of the individual masses still can easily be obtained 
by back-transforming each pair of $(m_c,\eta)$ samples. 
Fig.~\ref{gra:MassDensities} 
shows these two marginal PDFs combined into one plot.
\\ \indent
Analogously, other functions of the parameters can be derived and 
distributional features investigated; if e.g.\ one was interested in whether 
the masses differ `sig\-nif\-i\-cant\-ly' or are `almost equal', we can estimate:
$\prob(m_2 > 3m_1)=0.11\%$ or $\prob(m_2 < 1.5m_1)=4.84\%$. 
% > mean((dat$m2 > 3*dat$m1))
% [1] 0.001146362
% > mean((dat$m2 < 1.5*dat$m1))
% [1] 0.04843909
Fig.~\ref{gra:AlphaDensity} shows the posterior PDF of the 
logarithmic amplitude~$\alpha(m_1,m_2,d_E)$ (\ref{eqn:alpha}). 
Comparing it to the prior PDF you can see that, since it is significantly
above the reference points $x_U$ and $x_L$, the particular specification 
of the lower bound of the parameter space does not affect our conclusions. 
% \\ \indent
Table~\ref{tab:estimates} shows 
summary statistics of the
%numerical estimates of the  marginal 
posterior distributions of the inspiral's parameters.
\begin{table}[ht]
  \caption{Posterior estimates: Means, medians and 95\%~central posterior 
           intervals for several parameters.}
  \label{tab:estimates}
  %\begin{indented} \item[]
  \begin{center} \small
  \begin{tabular}{rlccccl}
    \br
    \multicolumn{2}{r}{parameter}  & mean   & median & 95\% c.p.i.       & true & unit\\ 
    \mr
    chirp mass & ($m_c$)           & 1.2161 & 1.2159 & $[1.2145,1.2186]$ & 1.2167 & $\sunmass$ \\
    mass ratio & ($\eta$)          & 0.2174 & 0.2162 & $[0.1987,0.2457]$ & 0.2222 &  \\ 
    coalescence time & ($t_c$)     & 84.6174 & 84.6174 & $[84.6160,84.6189]$ & 84.6167 & s    \\
    coalescence phase & ($\phi_0$) & \multicolumn{3}{c}{\emph{--- not meaningful ---}}& 0.0 & radian \\
    effective distance & ($d_E$)   & 26.28 & 25.99 & $[21.55,32.68]$ & 25.00 & Mpc \\[1ex]
    mass~1 & ($m_1$)               & 0.980 & 0.964 & $[0.876,1.229]$ & 1.0 & $\sunmass$ \\
    mass~2 & ($m_2$)               & 2.062 & 2.085 & $[1.600,2.327]$ & 2.0 & $\sunmass$ \\
    \br
  \end{tabular}
  \end{center}
  %\end{indented}
\end{table}
% >         mean     median       2.5%      97.5%
% mc   1.2160636  1.2159450  1.2144692  1.2186186
% eta  0.2173658  0.2161866  0.1986610  0.2457035
% tc  84.6173939 84.6174003 84.6159825 84.6188519
% phi  3.2203540  3.2891308  0.1538372  6.1284055
% dl  26.2786089 25.9878676 21.5453275 32.6797293
% m1   0.9804499  0.9635372  0.8757348  1.2290097
% m2   2.0617399  2.0845143  1.5999777  2.3270988

\section{Discussion}
We have developed an efficient method for examining gravitational wave
interferometer data for binary neutron star inspiral signals. 
%Our MCMC code searches for
%signals described by five parameters (two masses, effective distance, time, and
%phase at coalescence). This code can be applied at the end of a binary neutron
%star detection pipeline 
%\cite{AbbottEtAl2004b,MarionEtAl2003,AmicoEtAl2003},
%thereby producing parameter estimates and
%summary statistics.
Our MCMC code investigates the five parameters of an inspiral signal that is
described by its two masses, effective distance, time, and phase at coalescence.
This code can be applied at the end of a binary neutron star detection
pipeline \cite{AbbottEtAl2004b,MarionEtAl2003,AmicoEtAl2003}, 
thereby extracting detailed information about parameters
and related quantities.
\\ \indent
We are extending our MCMC research on binary neutron star inspiral signals. 
%Active research is in progress for more complex signals 
% from massive black hole systems, 
%
The power of MCMC methods is their ability to address complex signals
 with large parameter numbers. 
We see MCMC methods being of great use for even more complex 
 gravitational wave problems; 
 an example would be inspiral signals from black hole--black hole, 
 or black hole--neutron star systems,
 and binaries with significant amounts of spin for each mass. 
In order to address these signals higher order post-Newtonian terms 
 are necessary \cite{BlanchetEtAl2002}. 
A current and promising research effort of our group is in 
 incorporating these high order PN terms, 
 and to create an efficient  MCMC search routine 
 that will find all of the signal parameters.
Another on-going research project is in applying MCMC methods to the 
 multiple detector binary neutron star inspiral problem, 
 where sky position parameters can also be estimated. 
We expect that MCMC methods will prove useful 
 with more complex binary inspiral scenarios. 

\ack
This work was supported by the The Royal Society of New Zealand 
Marsden~Fund grant \mbox{UOA-204} and National Science Foundation grant \mbox{PHY-0244357}.

\section*{References}
  \bibliographystyle{unsrt}
  \bibliography{CQGinspiral}

\end{document}